# PRESSURE-THERMAL HYSTERESIS PHENOMENA IN BaTiO$_3$ UPON PHASE TRANSITION

Olga Mazur[1,a], Ken-ichi Tozaki[2], Leonid Stefanovich[1]

[1] Branch for Physics of Mining Processes of the M.S. Poliakov Institute of Geotechnical Mechanics of the National Academy of Sciences of Ukraine, Symferopolska st., 2a, Dnipro-5, 49600, Ukraine
[2] Department of Physics, Faculty of Education, Chiba University, Chiba-shi, Chiba, 263-8522, Japan
[a] Author to whom correspondence should be addressed: o.yu.mazur@gmail.com, +380509142572

## ABSTRACT

The phase transition into the ferroelectric phase in barium titanate occurs in many stages with the appearance of nonlinear phenomena. The mixed nature of the transition: displacive and order-disorder type, causes the occurrence of a thermal hysteresis, which span depends significantly on the pressure imposed on the sample. Theoretical calculations performed within the framework of Landau phenomenological theory showed a nonlinear dynamics of the thermal hysteresis with pressure. The existence of unusual relaxation phenomena was predicted in the pressure range of 130–145 MPa. Experimental and theoretical studies have shown the periods of slowing down of the domain structure upon the evolution process, the duration of which depends on the pressure and quenching temperature. It was assumed that relaxation processes near the tricritical point are similar to those observed near the Curie temperature. Theoretically, the parameters recommended for further experiments using high-pressure differential scanning calorimetry were calculated.

## GRAPHICAL ABSTRACT

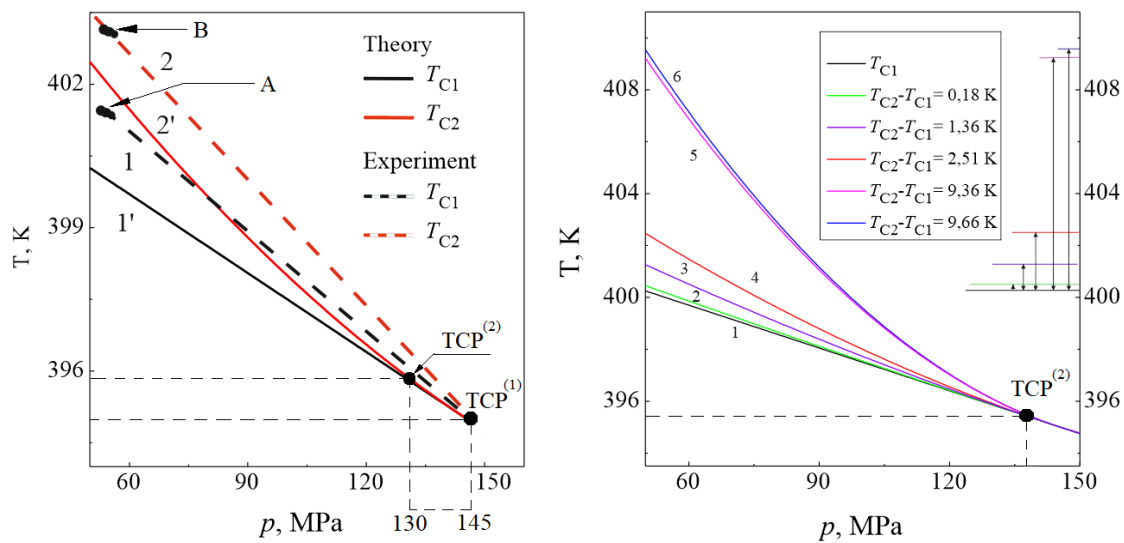

## 1. INTRODUCTION

Barium titanate (BaTiO$_3$) is a well-studied ferroelectric that undergoes several first-order structural phase transitions below Curie temperature $T_C$: in tetragonal ($T_C \sim$ 393–406 K), orthorhombic ($T_C \approx$ 280 K) and rhombohedral ($T_C \approx$ 185 K) phases. The thermodynamics of these phase transitions has been studied in sufficient detail [1–5]. But kinetic phenomena arising upon the primary formation of the ferroelectric domain structure require a deeper experimental study and theoretical interpretation.

The first-order phase transition between the paraelectric and ferroelectric phase in barium titanate occurs at different temperatures upon cooling $T_{C1}$ and heating $T_{C2}$ of the sample [6–13]. In BaTiO$_3$ single crystals, a thermal hysteresis $\Delta T = T_{C2} - T_{C1}$ is observed at all structural phase transitions: ~1–10 K in tetragonal; ~5.5 K in orthorhombic; and ~15 K in rhombohedral phases [8]. In polycrystalline ferroelectrics, the temperature span differs significantly, depending on the preparation conditions of the sample. The higher the annealing temperature of a ceramic BaTiO$_3$, the narrower the thermal hysteresis: ~19 K at 1273 K and ~12 K at 1473 K [6, 7]. Temperature dependent diffuse reflectance spectroscopy measurements showed the thermal hysteresis of ~2 K and ~40 K in powder and pellet BaTiO$_3$ samples respectively which suggests the large electronic disorder in the last one [13]. Structural and thermal disorder in the phase transition region determines the relationship between the thermal hysteresis and the appearance of metastable states [12]. This opens up the possibility of a kinetic influence on the sample, in particular, within the framework of strain engineering in order to improve the electronic qualities of materials for capacitors and energy storage devices [12–14].

Barium titanate has both ferroelectric and ferroelastic properties and is very sensitive to mechanical stress especially in the phase transition region where the shift of the Curie temperature with pressure $\partial T_C/\partial p$ is observed. The study of the pressure-induced thermal hysteresis in BaTiO$_3$ is of particular interest in order to establish the tricritical behaviour where the large value of dielectric permittivity, necessary for novel devices of the small size, is observed [1]. Previously, tricritical points in barium titanate were found under conditions of deep quenching at sufficiently high pressures: $T_{tcr}$ = 323 K [15]; $T_{tcr}$ = 291 K, $p_{tcr}$ = 3.4 GPa [16]; and $T_{tcr}$ = 130 K, $p_{tcr}$ = 6.5 GPa [4]. They are of a high fundamental interest, but cannot be used for practical purposes due to unsuitable values of control parameters to create the samples with desired features.

Recent studies predict the tricritical behaviour in BaTiO$_3$ upon nondeep quenching at much smaller pressure: $T_{tcr}$ = 380 K, $p_{tcr}$ = 0.33 GPa [17] and $T_{tcr}^{PTD}$ = 395 K, $p_{tcr}^{DSC}$ = 0.145 GPa [18]. The existence of the last point in BaTiO$_3$ single crystal was assumed by authors of this paper based on both experimental study and theoretical calculations. The results of differential scanning calorimetry (DSC), carried out under pressure in the range of 55-60 MPa, showed a gradual decrease in temperatures $T_{C1}$ and $T_{C2}$ with increasing pressure $p$. Linear extrapolation of the obtained points showed the convergence of temperatures $T_{C1}$ and $T_{C2}$ at pressure $p$ = 145.7 MPa. Based on the experimental data using the kinetic equations obtained in the framework of the Landau phenomenological theory, the pressure-temperature diagram (PTD) of the states of the domain structure was calculated, which showed the existence of a critical point at the values $T_{tcr}^{PTD}$ = 395 K and $p_{tcr}^{PTD}$ = 145 MPa [18]. But, as is known, the solution of stochastic equations essentially depends on the initial conditions, which cannot be controlled experimentally. Therefore, the accuracy of the calculations performed in [18] must be confirmed using other theoretical approaches.

This work is devoted to the study of the tricritical behaviour and thermal hysteresis span in BaTiO$_3$ single crystal proceeding from thermodynamic rather than kinetic considerations. One of the main tasks is to establish the correctness of the linear extrapolation of experimental results and the accuracy of theoretical calculations carried out in [18]. The mechanism of the metastable states accompanying the pressure-thermal hysteresis phenomenon will be studied. This research is aimed at developing a qualitative method for calculating control parameters (pressure and temperature) for the preparation of future experiments on the observation of critical phenomena upon phase transitions under pressure.

## 2. Dynamics of the thermal hysteresis with pressure

### 2.1. Theoretical consideration of the pressure-thermal hysteresis

Ferroelectric $BaTiO_3$ undergoes a first-order phase transition of a displacive type, but there are many evidences of its mixed character [8, 19, 20]. The occurrence of thermal hysteresis is associated precisely with the rotation of titanium Ti in octahedra $TiO_6$. But in temperature region up to 40 K above $T_C$ the order-disorder mechanism of the phase transition prevails [19]. At the same time, the Landau phenomenological theory [1, 18, 21, 22] describes the phase transition in $BaTiO_3$ more explicitly than microscopic models in conceptual understanding [23].

In temperature region $T < T_C$ both spontaneous polarization and spontaneous deformation emerge in $BaTiO_3$ crystal. But the last one only partially describes the decrease in symmetry arising upon phase transition. Therefore, the spontaneous polarization $\boldsymbol{P}_s$ is chosen as an order parameter, and the spontaneous deformation is a second-order effect. During the first-order phase transition the nonequilibrium thermodynamic potential requires the preservation of the sixth degree of the order parameter to ensure the stability of the ferroelectric state [23]:

$$F(p,T,P_z) = F_0(p,T) + \frac{1}{2}a(p,T)P_z^2 + \frac{1}{4}b(p)P_z^4 + \frac{1}{6}cP_z^6. \qquad (1)$$

Here $F_0(p,T)$ is a part of thermodynamic potential independent of the order parameter; $P_z$ is the projection of the spontaneous polarization vector on the polar axis of the crystal; $a(p,T) = \alpha(T - T_{C1}(p))$, where $T_{C1}(p)$ is the Curie temperature upon cooling of the sample; $b(p) = \beta(p - p_{tcr})$, where $p_{tcr}$ is the tricritical pressure; $c > 0$ is the constant.

The condition $\partial F/\partial P_z = 0$ implies the existence of five equilibrium values of the order parameter. All of them are valid only in the temperature range $T < T_{C2}(p)$, where $T_{C2}(p)$ is the Curie temperature upon heating of the sample:

$$T_{C2}(p) = T_{C1}(p) + \frac{b^2}{4ac}. \qquad (2)$$

In the symmetric phase ($T > T_{C2}(p)$) the free energy of the crystal (1) has only one minimum at $P_0 = 0$ (Fig. 1). Cooling the sample causes appearing of maxima $\pm P_{01}$ and minima $\pm P_{02}$. But as long as a minimum of $P_0$ exists, the crystal remains in the symmetric phase. Only when the sample is cooled to the temperature $T_{C1}(p)$, the state $P_0$ becomes unstable and the ferroelectric passes into the asymmetric phase with a jump in the order parameter $P_0^2(T_{C1}(p)) = -b/c$. Similarly, upon heating – the asymmetric phase can exist up to the temperature $T_{C2}(p)$: $P_0^2(T_{C2}(p)) = -b/(2c)$. Thus, temperatures $T_{C1}(p)$ and $T_{C2}(p)$ correspond to the boundaries where the paraelectric and ferroelectric phases can exist, respectively. The expression $\Delta T_C(p) = T_{C2}(p) - T_{C1}(p) = b^2/(4ac)$ characterizes the largest possible width of the thermal hysteresis [23].

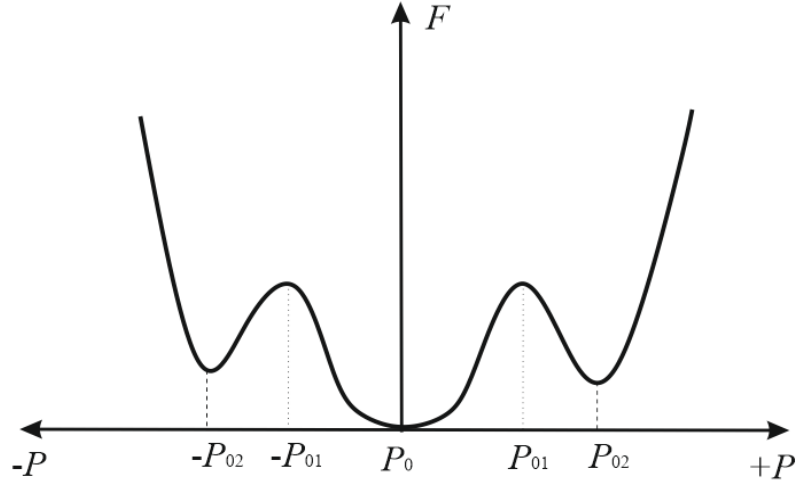

Fig.1. Dependence of the thermodynamic potential on the order parameter in the temperature region $T < T_{C2}(p)$, where the order parameter has five valid solutions

Both temperatures $T_{C1}(p)$ and $T_{C2}(p)$ and hence the span of the thermal hysteresis, $\Delta T_C(p)$ significantly depend on the pressure value $p$. For example, in $BaTiO_3$ it is 10 K and 2 K at pressures of 0.1 MPa and 20 GPa respectively [8]. The pressure dependence of the temperature $T_{C2}(p)$ is established in the parameter $b(p)$. The dependence of temperature $T_{C1}(p)$ on pressure can be written in terms of the ordering energy [24]:

$$T_{C1}(p) = T_C^{(0)}\left(1 - \frac{\sigma\varsigma}{\varepsilon_{at}}p - \frac{\gamma'\varsigma^2}{\varepsilon_{at}}p^2\right) \approx T_C^{(0)}\left(1 + \frac{p}{p_{int}}\right) = T_C^{(0)} + \gamma p . \qquad (3)$$

Here $T_C^{(0)}$ is the Curie temperature in the absence of pressure $p = 0$; $\varepsilon_{at}$ is the energy of interatomic interaction; $\sigma$ and $\gamma'$ are constants corresponding to the first and second derivatives of the interatomic interaction potential, respectively. Parameter $\varsigma = (1/3)\kappa r_0$ is related to the compressibility of the crystal $\kappa = (-1/V)(\partial V/\partial p)$, where $r_0$ is a radius of interatomic interaction. When the value of pressure imposed on the sample is much less than some internal pressure of the crystal $p_{int} = T_C^{(0)}/\sigma\varsigma$, i.e. when $p << |\sigma|/|\gamma'|\varsigma$, the expression (3) takes a form $T_{C1}(p) = T_C^{(0)} + \gamma p$. The internal pressure $p_{int}$ characterizes the repulsive force between atoms under stress. Parameter $\gamma = (\partial T_C/\partial p)$ is the baric coefficient, determining the shift of Curie temperature with pressure [24].

Expressions (2–3) show that in general case both temperatures $T_{C1}(p)$ and $T_{C2}(p)$ depend non-linearly on pressure. But in approximation $p << |\sigma|/|\gamma'|\varsigma$ the dependence $T_{C1}(p)$ becomes linear.

Theoretical analysis was provided in MatLab package using the following calculation parameters: $T_C^{(0)} = 403$ K [2], $\varepsilon_{at} = 0.011$ eV [25], $r_0 = 0.1$ nm [26], $\partial T_C/\partial p = -55$ K/GPa [27], $p_{tcr} = 145$ MPa [18]. Coefficients in the expansion (1) were chosen from different references for comparative analysis.

## 2.2. Comparison of theoretical and experimental results

The new tricritical point ($p_{tcr}^{DSC}$ = 145 MPa, $T_{tcr}^{DSC}$ = 395 K) was predicted experimentally in [18] by the linear extrapolation of Curie temperatures $T_{C1}(p)$ and $T_{C2}(p)$ upon cooling and heating respectively measured by DSC. But theoretical expressions (2–3) show that such extrapolation cannot be correct enough. Since the calculations were made at a relatively weak pressure $p$ = 55–60 MPa, the theoretical curve (curve 1') for the temperature $T_{C1}(p)$ is linear and is similar to the experimental curve 1 (Fig. 2). The theoretical curve 2', corresponding to the experimental extrapolation for the temperature $T_{C2}(p)$ (curve 2), exhibits a nonlinear character (Fig. 2). Theoretical curves 1' and 2' converge at pressure $p_{tcr}^{LG}$ = 130 MPa but not in the vicinity of the tricritical point $p_{tcr}^{DSC}$ = 145 MPa as was predicted in [18] (Fig. 2). Thus, the tricritical point calculated theoretically in this paper $TCP^{(2)}$ has values $p_{tcr}^{LG}$ = 130 MPa and $T_{tcr}^{LG}$ = 396 K (Fig. 2). The reasons for the significant difference between the current point $TCP^{(2)}$ and the previously one $TCP^{(1)}$ obtained experimentally and calculated from kinetic equations in [18] are described in Discussions. Further experimental research in the pressure range 130–145 MPa is of particular interest to establish the accurate value of the tricritical pressure.

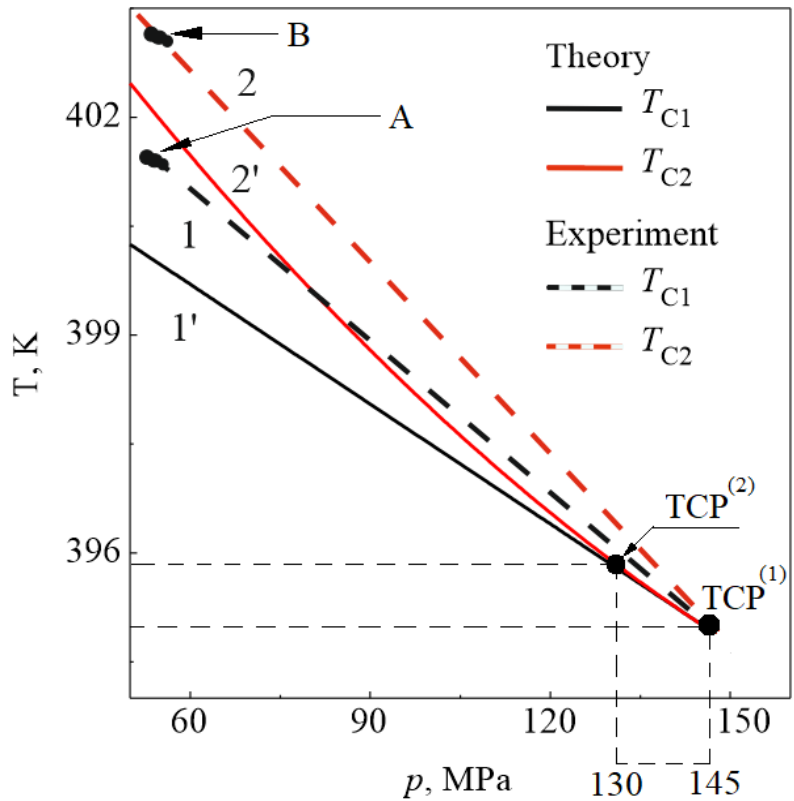

Fig. 2. Dependence of Curie temperatures upon cooling $T_{C1}(p)$ and heating $T_{C2}(p)$ of BaTiO$_3$ on pressure $p$ with indication of tricritical points: $TCP^{(1)}$ calculated by stochastic approach in [18] and by thermodynamic approach $TCP^{(2)}$. Black dots marked as A and B correspond to the temperature values $T_{C1}(p)$ and $T_{C2}(p)$ measured by DSC in the pressure range 55–60 MPa [18]. Dashed lines 1 and 2 are linear extrapolations of these points made in [18]. Solid lines 1' and 2' are theoretically calculated curves for temperatures $T_{C1}(p)$ and $T_{C2}(p)$ respectively from expressions (1–3)

The character of theoretical curves (1' and 2') for temperatures $T_{C1}(p)$ and $T_{C2}(p)$ significantly depends on the choice of the calculation parameters, in particular, the coefficients in the expansion (1). Curves in Fig. 2, were plotted using Landau coefficients in paper [28] and the results obtained are in good agreement with experimental data. Fig. 3 shows how much the curve for $T_{C2}(p)$ can differ depending on the Landau coefficients values. Curve 1 corresponds to the temperature $T_{C1}(p) = T_C^{(0)} + \gamma p$. Curves (2–6) correspond to temperatures $T_{C2}(p)$, where the Landau coefficients were taken from the papers [28, 30, 28, 31, 32] respectively. Curve 4 in Fig. 3 is the curve 2' in Fig. 2. It can be seen that with different initial data, the hysteresis can be both narrower (curves 2, 3 in Fig. 3) and wider (curves 5, 6 in Fig. 3).

Despite the large scatter of the calculated values for $T_{C2}(p)$, the obtained curves qualitatively repeat the conclusions mentioned above. First, regardless of the width of thermal hysteresis, all $T_{C2}(p)$ curves converge with the $T_{C1}(p)$ curve at pressures $p \geq 130$ MPa. Second, curves 5 and 6 in Fig. 3 clearly illustrate the nonlinear nature of the dependence of temperature $T_{C2}(p)$ on pressure $p$.

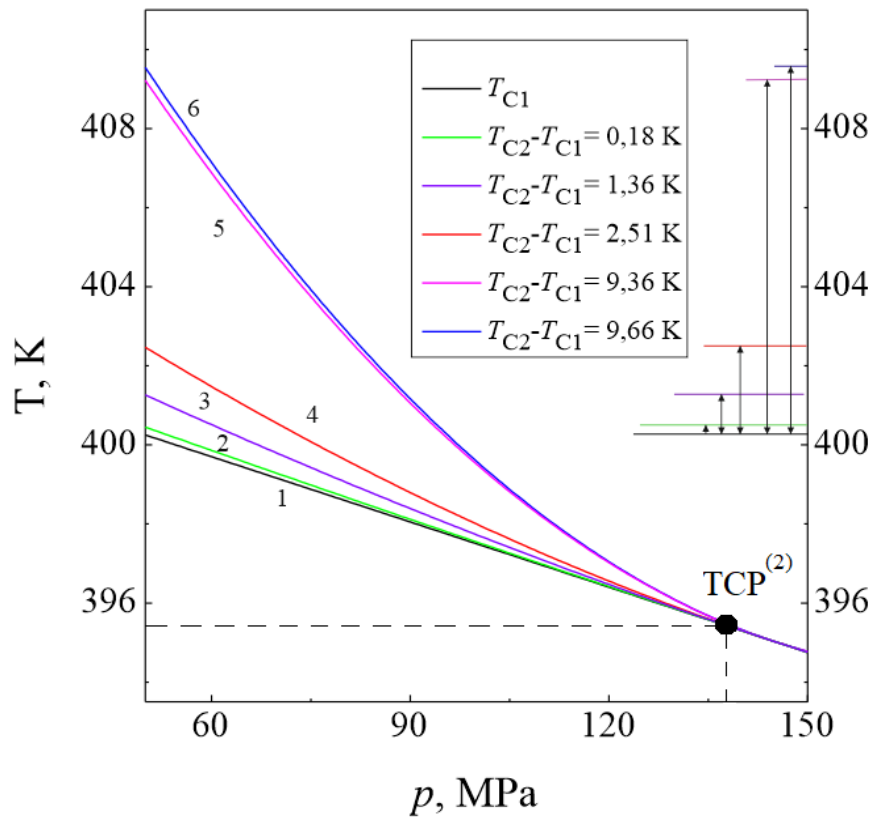

Fig. 3. Theoretical curves for temperatures $T_{C1}(p)$ (curve 1) and $T_{C2}(p)$ (curves 2–6), depending on pressure $p$ calculated with different Landau coefficients from papers [33, 34, 32, 35, 36] respectively. The corresponding thermal hysteresis widths are indicated in the legend

## 3. Formation of metastable states upon cooling/heating the sample

### 3.1. Experimental details.

#### 3.1.1. Equipment and sample

The new DSC experiment of the pressure-induced phase transition in $BaTiO_3$ was carried out using the equipment and parameters described in detail in [18]. It is known that the thermal hysteresis significantly depends on both experimental conditions and parameters of the sample. For example, it was shown, that in $BaTiO_3$ films with a thickness of ≥ 100 nm it was 5 K and more, while at a thickness of ~ 40 nm it completely disappeared [33]. In our experiments, we used a sample with a scale of 4×4×1.2 $mm^3$ prepared as written in [18]. The sample was heated and cooled at different pressures, which allowed to establish a decrease in the width of thermal hysteresis with subsequent extrapolation to the tricritical point.

#### 3.1.2. Reproducibility

For the quantitative examination of the thermal hysteresis as an irreversible process, it is essential to gain experimental reproducibility. Repetition of the cooling-heating loop with a constant speed used in [18] is an explicit experimental condition to obtain good reproducibility. Figure 4 shows the heat release upon cooling and heat absorption upon heating during the phase transition as a function of time. The measurement interval is 2 s, and the temperature sweep speed is ±500 μK/s. The lower limit of measurement interval is bounded by the response time of the apparatus.

The plots are superposed for 26 repetitions of the temperature scanning loop by adjusting the time axis for each scan so that the curve overlaps at the latter relaxation part (Fig. 4). The superposed plots of the exotherm and endotherm represent a single curve, respectively, and the total amount of heat transfer indicated by the area is the same. These results confirm the excellent experimental reproducibility of this nonequilibrium process. By superimposing the repeated measurement data with a technique using the isomorphicity at the relaxation part, we know the thermogram shape with a shorter time resolution than the measurement interval.

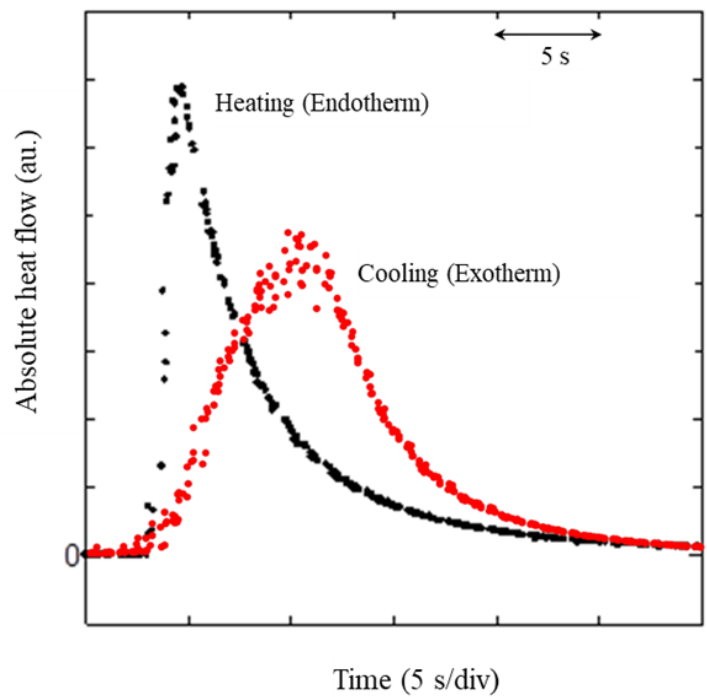

Fig. 4. Superposed thermogram for 26 cooling-heating loops at pressure range 55–60.2 MPa. The time axis is adjusted for each scan so that the curve overlaps at the latter relaxation part

The exotherm and endotherm traces can be divided into two stages: the former beginning stage and the latter relaxation one respectively. The shape of the thermogram in the beginning part is made by heat transformation within the crystal, and that in the latter part shows how the heat transfers between the crystal and the surroundings. The amount of heat transfer between the crystal and the surroundings is proportional to the temperature difference and is independent of the temperature sweep direction. Thus, the shape of the thermogram at the latter part is the same for both cooling and heating. Comparison of the thermograms shows that the heat absorption upon heating completes in a shorter time than the heat release upon cooling. This fact agrees that the anomaly at this phase transition is observed ordinarily upon cooling and not heating, as in Forsbergh birefringence patterns studies using a polarizing microscopy [34].

Furthermore, the beginning part of both thermograms seems not smooth but stepwise. This behaviour suggests that unknown structure affects this irreversible transition process.

### 3.1.3. Further exploration

The rate of cooling/heating $v > 6$ mK/s can slow down/accelerate the beginning of the phase transition [8]. For a further detailed examination of the irreversible transition process, the experiment should be conducted at a possible slow cooling rate. We discuss below the experimental results at a cooling rate $v = 1$ µK/s, which cooling speed is 500 times slower than mentioned above [18]. Even at such low speeds, the ferroelectric crystal at the quenched and metastable state does not go to an equilibrium at once and remains susceptible to pressure. The arising of the first ferroelectric domains occurs in $10^{-12} - 10^{-6}$ s [35]. Therefore, the observation of the ferroelectric phase formation at early stage is possible at the slow cooling of the sample from the quenched quasi-equilibrium state.

### 3.2. Intermediate states upon phase transition

The new DSC experiment was made in the vicinity of phase transition using the equipment and samples as in [18] to study metastable phenomena accompanying the pressure-thermal hysteresis. The BaTiO$_3$ single crystal was quenched with a cooling rate $v = 1$ µK/s in the vicinity of Curie temperature ($T_C - T < 1$ K) at pressure $p = 0.101$ MPa. Such low pressure and closeness to the phase transition point allow observing the "extended-in-time" phase transition with the formation of kinetic intermediate states.

It was shown that the temperature range for the realization of the metastable phase upon heating is narrower than upon cooling (Fig. 5). The magnitude of total entropy change due to the transition agrees with that in CsPbCl3 [36]. At low pressure $p = 0.101$ MPa in the vicinity of $T_C$ ($T_1 = 402.278$ K), a heat flow peak is observed, indicating the onset of a metastable phase (Fig. 6a). The system remains in this state for a quite long time $t_1 = 17218$ s, after what the second heat flow peak occurs at the temperature $T_2 = 402.261$ K and a coherent hybrid structure (CHS) is formed, consisting of domains of tetragonal and monoclinic phases (Fig. 6) [2]. These heat flow peaks can be estimated within the framework of Landau theory (Fig. 6b) from the relation between the heat capacity and Landau coefficients [37]. The fluctuation part (Fig. 6c) of this heat can be predicted from the theory of fluctuation phenomena [5]. The theoretical curves (Fig. 6b) confirm that the heat flow peak for $T_1 = 402.278$ K (curve 1) is lower than for $T_2 = 402.261$ K (curve 2), as in the experimental plot (Fig. 6a). Although the obtained theoretical curves qualitatively repeat the behaviour of the BaTiO$_3$ crystal observed experimentally [2], they give a reliable description only at very low pressure ($p = 0.101$ MPa) or in its absence. The strong anisotropy of the soft mode dispersion leads to a significant difference in the obtained values and should be included in theoretical expressions.

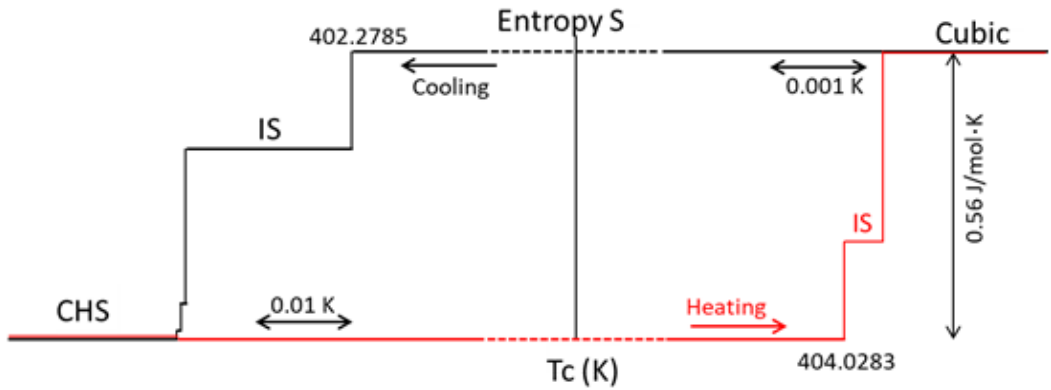

Fig. 5. Entropy-temperature hysteresis in BaTiO$_3$ upon cooling and heating with indication of thermal ranges for existence of an intermediate state (IS) and the formation of a coherent hybrid structure (CHS). Cooling trace is estimated from ref. [18]. The cooling rate is 1 μK/s, and heating rate is 10 μK/s.

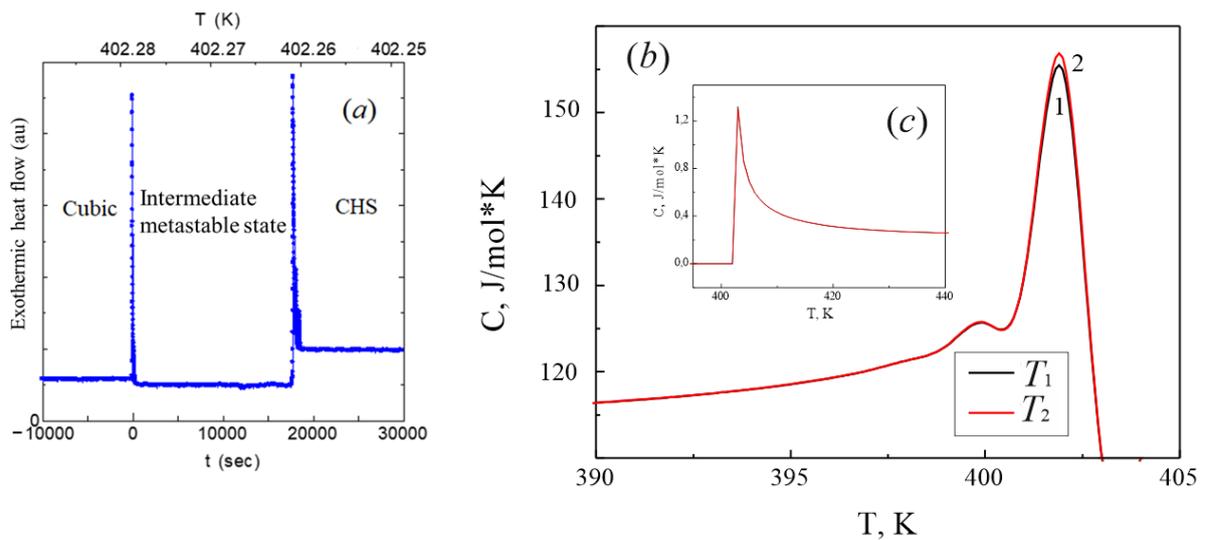

Fig. 6(a–c). Two-step phase transition in BaTiO$_3$ upon cooling: (*a*) – experimental heat flow peaks with indication of the relaxation time of intermediate phase [2]; (*b*) – theoretically obtained curves of the heat capacity for temperatures $T_1 = 402.278$ K (curve 1) and $T_2 = 402.261$ K (curve 2); (*c*) – fluctuation part of the heat capacity.

Numerous studies indicate that both thermal hysteresis and metastable phases are closely related to the transformation of domains [8, 38–40]. The change in the properties of crystals under pressure occurs as a result of polarization switching and dynamics of domain walls [41]. Therefore, the dynamics of a domain structure quenched near the Curie temperature ($T_C - T < 10$ K) under a hydrostatic pressure was studied within the framework of the previously developed kinetic model [18].

Evolutionary curves were plotted for the average polarization $\bar{\pi}$ (Fig. 7) and allowed to estimate the duration of slowing down of the system in the metastable phase depending on quenching temperature (Fig. 7a) and pressure (Fig. 7b,c). They showed that a metastable phase is formed with a pronounced asymmetry in the volume fractions of differently directed domains [18]. The ferroelectric in this state does not experience any dynamics, and this intermediate stage is realized in a wide temperature range (Fig. 7a). But in the case of a weak external pressure $p = 0.101$ MPa it exists for a sufficiently long time only in the proximity of Curie temperature $T_C - T < 1$ K (curves 1, 2 in Fig. 7a). As the quenching temperature increases, the time of existence of the intermediate state gradually decreases (curves 3, 4 in Fig. 7a).

The duration of the metastable phase upon quenching the system to the temperature $T_1 = 402.278$ K is theoretically estimated as ~9000 s (curve 1 in Fig. 7a) compare to the experimental value $t_1 = 17218$ s (Fig. 6a). It is related with the limited consideration of only 180°-domains by theoretical model of one-component order parameter. The extension of this approach for the three-component order parameter in the future will significantly reduce the calculation error and allow the establishment of different domain walls.

The value of the tricritical pressure $p_{tcr}^{DSC,PTD} = 145$ MPa obtained in [18] requires additional verification, but not the value of the tricritical temperature $T_{tcr}^{PTD} = 395$ K, which is confirmed by the DSC measurements $T_{tcr}^{DSC} = 394.3$ K [18] and current study $T_{tcr}^{LG} = 396$ K (Fig. 2). Therefore, evolutionary curves were plotted for the case when the sample is quenched to this temperature in order to study kinetic phenomena depending on pressure value (Fig. 7b,c). It was shown that upon such quenching a low pressure does not cause any special relaxation anomalies in the system, and metastable phases can exist only for a very short time (curve 4 in Fig. 7a and curve 1 in Fig. 7b,c). But with increasing pressure, the duration of the kinetic deceleration of the system raises (curves 2, 3 in Fig. 7b,c) and increases to the maximum when approaching the value of the pressure $p = 145$ MPa. But already at a pressure $p = 130$ MPa, the duration of the intermediate phase reaches ~10000 s (curve 3 in Fig. 7b). Therefore, further DSC experiments in the pressure range 130–145 MPa seem very promising, since they can show the presence of many relaxation anomalies occurring very slowly.

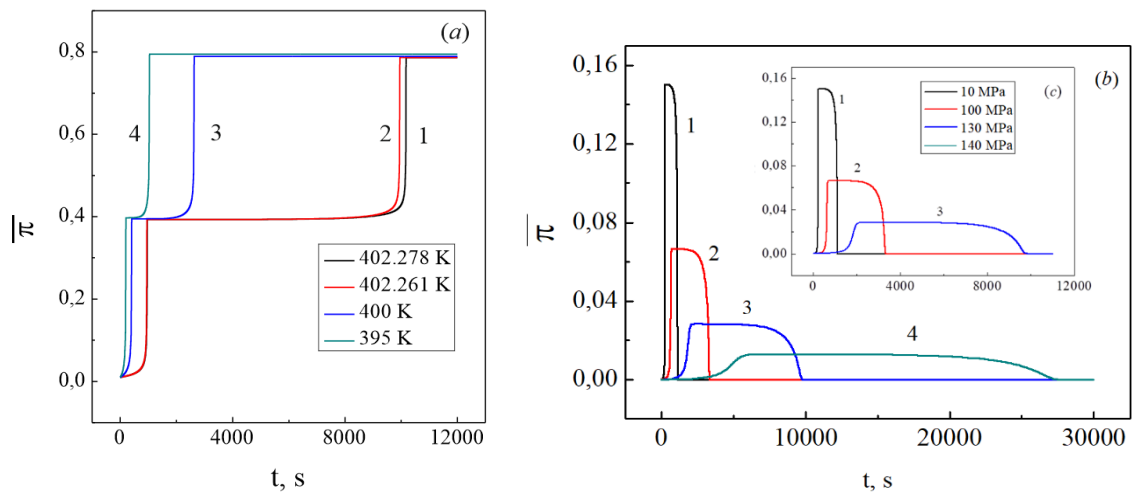

Fig. 7(a–c). (a) – Evolution curves for average polarization $\bar{\pi}$ at constant pressure $p = 0.101$ MPa depending on quenching temperature: 402.278 K; 402.261 K; 400 K; 395 K for curves 1–4 respectively; (b) – evolution curves for the average polarization $\bar{\pi}$ upon quenching in the vicinity of the temperature $T = 395$ K depending on pressure: 10 MPa; 100 MPa; 130 MPa; 140 MPa for curves 1–4 respectively; (c) – enlargement plot of the Fig. 7b showing the relaxation plateau for curves 1–3

Within the framework of the current kinetic model, the rate of domain enlargement and relaxation times were estimated, depending on the quenching temperature (Fig. 8*a*) and pressure (Fig. 8*b*) respectively. The time dependence of the average domain size $\bar{R}$ has a root law [42]. Near the Curie temperature $T_C$ all processes in the system are very slow, including the growth of domains. In this region the nucleation of a large number of new small-sized domains prevails, and a high degree of heterogeneity in the system remains (curve 1 in Fig. 8*a*). Deep quenching at the same rate occurs longer, so by the relaxation begins, rather large domain regions have already been formed in the system (curve 2 in Fig. 8*a*), and the overall structure inhomogeneity is reduced.

The relaxation times of the domain structure were calculated depending on the pressure and quenching temperature (Fig. 8*b*). It was shown that in the vicinity of the Curie temperature, the relaxation of the system occurs gradually and very slowly (curve 1, Fig. 8*b*), and the relaxation time weakly depends on pressure. As the quenching temperature increases, the domain structure forms faster (curves 2 and 3, Fig. 8*b*), and the relaxation time shows nonmonotonic dependence on pressure. In this case, there are pressure ranges where the relaxation time of the system is practically independent of the pressure. In particular, upon quenching at 400 K and 398 K the relaxation time changes very slightly at pressures up to 40 MPa, and up to 70 respectively. This indicates that at the given quenching temperatures, the domain structure develops rapidly, and the pressure only stimulates numerous domain rearrangements.

In the case of sufficiently deep quenching $T = 395$ K, the domain structure is formed very fast at pressures up to 110 MPa, and the relaxation time is almost independent on pressure. However, starting from 130 MPa and up to 145 MPa, the relaxation time drastically increases and respectively the domain structure evolution slows down considerably (curve 4, Fig. 8*b*). Curves 1 and 4 in Fig. 8*b* show a similar behaviour which can indicate that a relatively large pressure near tricritical point causes the same relaxation phenomena as were observed at weak pressure near the Curie temperature.

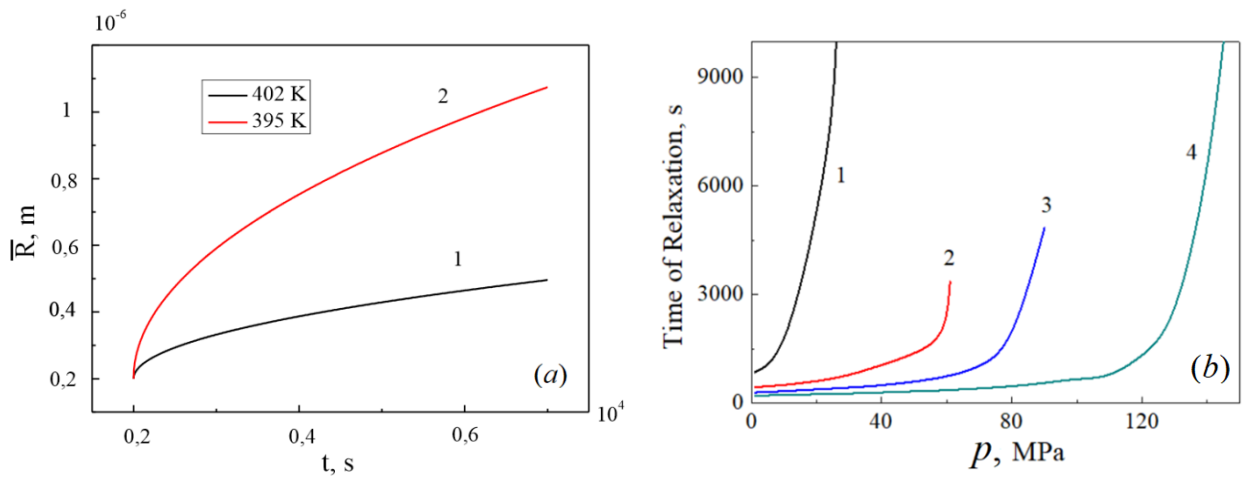

Fig. 8(a,b). (*a*) – Time dependence of the average domain size in BaTiO$_3$ crystal quenched to the temperatures 402 K and 395 K for curves 1, 2 respectively; (*b*) – theoretical dependence of the relaxation time of the domain structure in BaTiO$_3$ on the pressure and quenching temperature: 402 K; 400K; 398K; 395 K for curves 1–4 respectively

## 4. DISCUSSIONS AND CONCLUSION

Hysteresis phenomena arising upon phase transition under the influence of an external electric field have been studied quite well both theoretically and experimentally [21, 43–47] compare to the thermal and pressure-thermal hysteresis. This is due to the significant dependence of the latter, both on the sample parameters (structure, scale, defectiveness, prehistory, etc.), and on the experimental control parameters of phase transition (external pressure, quenching temperature and rate).

The difference of Curie temperatures $T_{C1}(p)$ and $T_{C2}(p)$ upon cooling and heating respectively which gradually decreases with increasing pressure and quenching temperature was found experimentally in [18]. As a result of linear extrapolation of these data the existence of tricritical point $TCP^{(1)}$ at values $p_{tcr}^{DSC} = 145.7$ MPa and $T_{tcr}^{DSC} = 394.3$ K was predicted. Based on experimentally obtained value of tricritical pressure $p_{tcr}^{DSC} = 145.7$ MPa the phase diagram was created from the stochastic approach to predict the state of a thermodynamically stable domain structure depending on pressure and quenching temperature. There was found a point at $p_{tcr}^{PTD} = 145$ MPa and $T_{tcr}^{PTD} = 395$ K (at certain initial parameters) where the single-domain state does not exist anymore. Previous research on ferroelectric phase transitions under pressure showed that a single-domain state is realized predominantly upon the first-order phase transition and almost not observed upon the second-order [48, 49]. Thus it was assumed that the point this point characterizes the change of the character of phase transition from the first-order to the second-order and can be associated with the point $TCP^{(1)}$. But the solution of stochastic equations significantly depends on the choice of initial parameters that cannot be controlled experimentally: initial correlation radius $r_c(0)$, initial average polarization $\bar{\pi}_0$ and its dispersion $D_0$. Varying these parameters and setting different degree of initial inhomogeneity of the system, as well as the initial size of domains, can lead to a significant shift of the point $TCP^{(1)}$. In addition, the theoretical model in [18] did not affect the study of the width and dynamics of the pressure-thermal hysteresis.

The current work presents a critical evaluation of the point $TCP^{(1)}$ obtained in [18]. To avoid the influence of stochastic errors, the calculation was carried out not from stochastic, but from the thermodynamic considerations (expressions 1–3). The convergence of temperatures $T_{C1}(p)$ and $T_{C2}(p)$ was shown to occur at pressure $p_{tcr}^{LG} = 130$ MPa (Fig. 2, 3), which is 15 MPa less than the previously calculated value [18]. At the same time, the span of thermal hysteresis $\Delta T_C(p) = 2.51$ K (at $p = 55$ MPa) (Fig. 3) calculated in this research agrees well with the current experimental results (Fig. 2) and known data. In particular, thermal hysteresis $\Delta T_C \approx 2$ K was experimentally observed near $T_C^{(0)} = 402$ K in [10]. The calculation of the relaxation time of the system shows the slow occurrence of relaxation processes near the phase transition point and upon deep quenching ($T_C - T = 8$ K) at pressures 130–145 MPa (Fig. 8b). Therefore, it can be assumed that the exact value of the tricritical pressure is located in this range. According to the results obtained the pressure range 130–145 MPa will be chosen for the preparation of the future experiment to accurately establish the value of the tricritical point and determine the theoretical approach more suitable for describing kinetic phenomena upon phase transition under pressure.

The phase transition in $BaTiO_3$ occurs unevenly. As shown in [8], at cooling/heating rates $v < 1-3$ mK/s, both transitions go on as a series of steps in the temperature range 1–3 K. Recent pressure experiments confirmed that the phase transition is carried out in an order-disorder type two-step structural change upon cooling and heating both in bulk samples [2, 38] and in thin films of $BaTiO_3$ [33]. In this case, an intermediate state appears, after which a hybrid structure is formed, consisting of tetragonal (major) and monoclinic (minor) domains [2]. The existence of metastable phases during phase transitions in $BaTiO_3$ was predicted in many studies [8, 20, 34, 50]. The character and duration of intermediate states essentially depend both on the factors of external influences and the type of transition.

The mechanisms of the existence of metastable phases are quite complex. In [20], using atomistic simulation, it was shown that in $BaTiO_3$ the most important role in this process is played by the dynamics of polar clusters in the paraelectric phase, as precursors to the phase transition. The discontinuous nature of the transition leads to the formation of an intermediate metastable state containing regions of both the still symmetric high-temperature phase and the already formed asymmetric low-temperature phase. The study of the domain dynamics in the molecular scale showed that the mixing of paraelectric and ferroelectric configurations causes a pronounced imbalance between oppositely directed ferroelectric domains [20]. These statements correlate well with our results on the existence of a long-term metastable phase with a pronounced asymmetry in the volume fractions of domains of different signs (Fig. 7).

The current study showed a significant dependence of relaxation processes in barium titanate characterizing the domain structure formation on pressure. Understanding of kinetic mechanisms of phase transformation under external pressure opens up possibilities for directed control of ferroelectric ordering and creation the samples with given parameters. Further studies of the kinetics in the vicinity of Curie temperature and tricritical point are of both fundamental and practical interest.

## DECLARATION OF INTERESTS

The authors declare that they have no known competing financial interests or personal relationships that could have appeared to influence the work reported in this paper.

## ACKNOWLEDGMENTS

This work was funded by the project № 9918 of the IEEE program "Magnetism for Ukraine 2022" supported by the Science and Technology Center in Ukraine (STCU) together with the Institute of Magnetism NASU and MESU, on behalf of the IEEE Magnetic Society.

O. Mazur thanks the Wolfgang Pauli Institute for providing support in "Pauli Ukraine Project" within the WPI Thematic Program "Mathematics-Magnetism-Materials (2021/2022)".

K. Tozaki thanks Dr. Y. Yoshimura for his valuable discussions.

## DATA AVAILABILITY

The data that support the findings of this study are available from the corresponding author upon reasonable request.